\documentclass[11pt,twoside]{article}

\usepackage{asp2006}
\usepackage{graphicx}
\usepackage{lscape}

\newcommand{\redacted}[1]{}

\begin{document}
\title{Morphology and dynamics of photospheric and chromospheric magnetic fields}   
\author{Friedrich W\"oger}   
\affil{National Solar Observatory, PO Box 62, Sunspot, NM 88349, USA}    
\author{Sven Wedemeyer--B\"ohm}   
\affil{Institute of Theoretical Astrophysics, University of Oslo, Postboks 1029 Blindern, N-0315 Oslo, Norway}    
\author{Thomas Rimmele}   
\affil{National Solar Observatory, PO Box 62, Sunspot, NM 88349, USA}    

\begin{abstract} 
We use joint observations obtained with the Hinode space observatory and the Interferometric Bidimensional Spectrometer (IBIS) installed at the DST of the NSO/SP to investigate the morphology and dynamics of (a)~non-magnetic and (b)~magnetic regions in the fluctosphere.
In inter-network regions with no significant magnetic flux contributions above the detection limit of IBIS, we find intensity structures with similar characteristics as those seen in numerical simulations by \citet{wedemeyer2008b}.
The magnetic flux elements in the network are stable and seem to resemble the spatially extended counterparts to the underlying photospheric magnetic elements.
We will explain some of the difficulties in deriving the magnetic field vector from observations of the fluctosphere.
\end{abstract}


\section{Introduction}
In the past years, the chromosphere and its magnetic structure has been found to be a key ingredient to progress in understanding the solar corona.
Yet, little is known of the properties of the chromosphere on small scales.

Recent high-resolution observations suggest that in the quiet Sun there exists a weak-field domain below the classical ``canopy'' \citep{woeger2006}.
The prominent magnetic fields of the ``canopy'' are major constituents of the chromosphere in the stricter sense as it is seen in the H$\alpha$ line.
The weak-field domain below is referred to as ``fluctosphere'' hereafter -- a term coined by \citet{wedemeyer2008a}.
The fluctosphere in numerical LTE radiation HD models is
generated by the interaction of (acoustic) shock waves, which are excited in the photosphere and propagate upwards into the layer above.
At a height range in the fluctosphere around 1000\,km above $\tau=1$ this
results in an apparent temperature pattern that has similarity to reversed
granulation, yet changes with typical timescales between 20--30\,s
\citep{wedemeyer2004}. Synthetic spectra from numerical simulations
\citep{wedemeyer2008b} allows us to directly compare the models with
observations of this layer in order to study its morphology and dynamics.

As very little is known about the fluctospheric magnetic fields, we present and interpret for the first time high spatially resolved spectro-polarimetric measurements in the fluctosphere above strong photospheric magnetic features.
\vspace{-0.25cm}

\begin{figure}[t]
\centering
\includegraphics[width=0.7\textwidth]{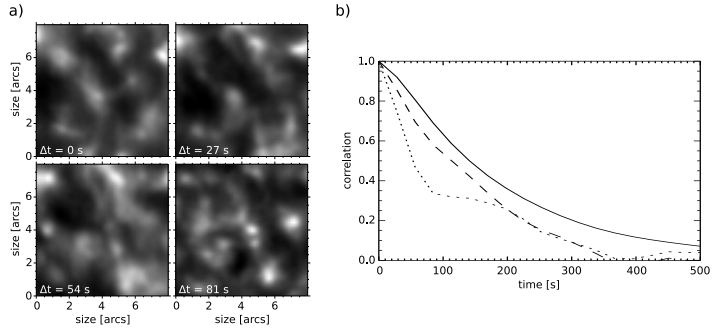}
\caption{a) Temporally consecutive subframes of a weak-field region in the Ca II infrared line core. All images are equally scaled. b) Auto-correlation functions of the two-dimensional intensity of continuum (solid), line wing at 854.2457\,nm (dashed), and of the line core at 854.2158\,nm (dotted).}
\label{fig:corr}
\end{figure}
\section{Observations}
We observed a quiet Sun region located in a coronal hole near disc center on May 26, 2007 using the Ca II infrared line at 854.2\,nm.
The observations were carried out in coordination with Hinode to simultaneously obtain G-band images.
Several persistant G-band bright points are visible in the field of view, forming a network element.

The IBIS instrument consists of two channels, a narrowband channel with two tunable Fabry-Perot interferometers and a broadband channel for reference \citep{cavallini2006}.
The broadband images of IBIS were calibrated with the standard procedure of dark subtraction and gain table application and reconstructed using the speckle image reconstruction algorithm modified for high-order adaptive optics corrected data detailed in \citet{woeger2008}.
The reconstructed broadband images were used to destretch the narrowband images of IBIS.
Finally, the Hinode G-band images were aligned to the broadband reconstructions.

The IBIS narrowband channel was set up in spectro-polarimetric mode to scan the line core of the Ca II line in 2 dimensions using six modulation states and an exposure time of 35\,ms.
The line was scanned with 17 wavelength steps using a step width in wavelength corresponding to 4.3\,pm with a full width at half maximum transmission of 4.6\,pm.
This procedure resulted in a overall cadence of approximately 27\,s over the 26.5 minutes observation sequence.
A subregion of the data set displayed for the Ca II infrared line core is shown in Fig.~\ref{fig:corr}a for a small part of the time sequence.

\begin{figure}[t]
\centering
\includegraphics[width=0.7\textwidth]{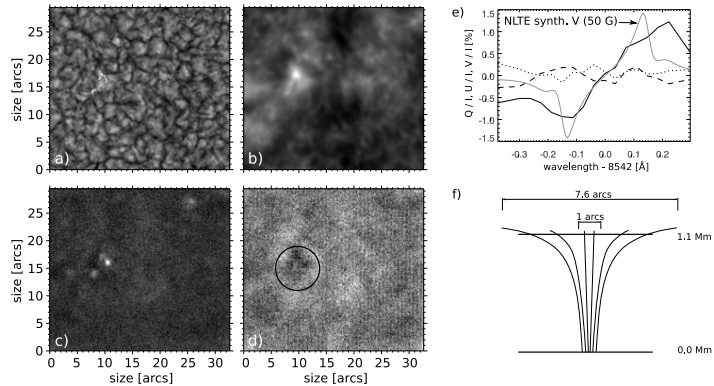}
\caption{Temporal averages (26.5\,min) of a) G-band intensity, b) Ca II line core intensity, c) total circular polarisation, d) total linear polarisation, and e) Stokes profiles of the strong magnetic features located at [11,16]\,arcs (black solid: Stokes V/I; dashed: Stokes U/I; dotted: Stokes Q/I). The ring inscribed in d) has a diameter of $\sim$7.6 \,arcs. f) Sketch of an idealized magnetic field geometry that could explain the observed signals.}
\label{fig:mag}
\end{figure}
\section{Analysis}
\noindent{\bf a) Properties of the intensity pattern in inter-network regions}\\
In our analysis, we have measured the typical time scales with which the two-dimensional structures measured at different positions in the Ca II infrared line evolve.
Using the two-dimensional auto-correlation function, the $1/e$ time scale was computed for continuum, line wing and line core of the Ca II infrared line in weak-field regions.
We obtain similar results as in \citet{leenaarts2005} and \citet{woeger2006}, who used the Ca II H and K lines, respectively, with a correlation time of 176 seconds for granulation, about 152 seconds for reversed granulation and 59 seconds for the pattern in the fluctosphere (see Fig.~\ref{fig:corr}~b)).
Nevertheless, in comparison to the results of these works, the spectral bandwidth of the presented observations is narrower by an order of magnitude, which has implications on the height origin of the specific intensity.
Models of the Ca II infrared contribution function predict significant contributions from photospheric layers when observing with broad filters \citep{uitenbroek2006}, demanding the narrowest possible filters to attribute measured intensity to the fluctosphere and chromosphere.
The observed data indicate the existence of a two-dimensional pattern originating from heights around 1000-1500\,km that evolves faster than e.g. reversed granulation.
This pattern is likely the intensity signature of the temperature pattern seen in the radiation hydrodynamic simulations of \citet{wedemeyer2004}.

\noindent{\bf b) Topology of the magnetic network elements}\\
We have analyzed the field structure of strong magnetic features extending from photosphere to fluctosphere by averaging the total linear and circular polarization signals over the time sequence, as the signal-to-noise ratio was insufficient for an analysis of single scans.
The result of this averaging is shown in Fig.~\ref{fig:mag}a--d.
The observations, acquired in a coronal hole region, show a distinct signal in the total circular polarization signal with a diameter of 1\,arcs at the locations of the magnetic features in the photosphere which are best seen in G-band images.
These features remain at the same location for the duration of the observation and -- while their movement is dominated by the convective motion -- do not travel over long distances.
Thus, in the averaged G-band images the location of the magnetic features are clearly visible (Fig.~\ref{fig:mag}a).
The temporal averaging also reveals a ring in total linear polarization with a radius of 7.6\,arcs, possibly indicating a funnel-like structure in the fluctospheric layers.
We intend to analyze the temporal behavior of the magnetic fields in the fluctosphere in the future.

The Stokes parameters averaged over the time sequence are displayed in Fig.~\ref{fig:mag}e.
When compared to a NLTE model of the Ca II infrared line at 854.2\,nm \citep[RHSC3D,][]{uitenbroek2000}, assumed for disk center geometry, the observed Stokes~V/I signal compares well to a signal generated by a magnetic field strength of 50\,Gauss.

\section{Discussion and Conclusion}
We present spectro-polarimetric observations of the solar atmosphere in a coronal hole with focus on the fluctosphere.
We find that (a) inter-network regions in the fluctosphere display an intensity pattern that shows structures of similar spatial scales as reversed granulation, yet is evolving on much faster time scales. We interpret this pattern as the intensity signature of propagating, interacting shock waves as predicted by three-dimensional LTE radiation hydrodynamic models of the fluctosphere.
(b) The strong magnetic features visible in our observations possibly form a funnel-like structure in the fluctosphere that is likely extending into the chromosphere. The Stokes V/I signal in the strongest magnetic feature visible compares well to a signal created by a field strength of 50\,Gauss in NLTE simulations \citep[RHSC3D,][]{uitenbroek2000}.

Measurements of the magnetic fields in the fluctosphere and chromosphere are important when trying to gather a deeper understanding of the chromospheric energy balance.
As the magnetic structures in these layers are likely to be small, ultimately telescopes with larger apertures are needed.
The difficulties of the observation of weak magnetic fields in the fluctosphere and chromosphere are only surpassed by the interpretation of such data.
In general, the height origins of the Stokes signals are unclear and do not
only depend on contribution function but also on the height distribution of the magnetic field strength.

\acknowledgements 
The authors thank Alexandra Tritschler, Han Uitenbroek and Kevin Reardon for helpful discussions.


\end{document}